# Modeling the radial distribution of blue stragglers in M3

*Steinn Sigurdsson*[1], *Melvyn B. Davies*[2] *and Michael Bolte*[1]

[1]Lick Observatory, University of California, Santa Cruz, CA 95064

[2]103–33, California Institute of Technology, Pasadena, CA 91125

astro-ph/9406004  1 Jun 1994

submitted to the *Astrophysical Journal, Letters*

This research was supported in part by NASA grant NAGW–2422,

the NSF under Grants AST 90–18526 and ASC 93-18185

and an R.C. Tolman Fellowship awarded to M.B.D.

# Modeling the radial distribution of blue stragglers in M3

*Steinn Sigurdsson*[1], *Melvyn B. Davies*[2] *and Michael Bolte*[1]

[1]Lick Observatory, University of California, Santa Cruz, CA 95064

[2]103–33, California Institute of Technology, Pasadena, CA 91125

## ABSTRACT

Recent surveys of the blue straggler (BS) population in the galactic globular cluster M3 (NGC 5272) give the first complete characterization of the number density of BSs as a function of radius over an entire globular cluster. The BSs in M3 are overabundant at large radii, and significantly underabundant at intermediate radii. Here we present the result of a simulation of the dynamical evolution of a population of BSs in a multi–mass model of M3. Assuming the BSs were formed in the core through binary interactions (Hut & Verbunt 1983, Leonard 1989, Sigurdsson & Phinney 1993, Hut *et al.* 1992, Davies, Benz and Hills 1994), and given some very general assumptions about the recoil that occurs during stellar mergers in interacting binaries; we find an excellent fit to the observed radial distribution of BSs, suggesting strongly that most of the BSs in M3 were formed through binary collisions in the core.

May 1994

submitted to the *Astrophysical Journal, Letters*

---

This research was supported in part by NASA grant NAGW–2422,

the NSF under Grants AST 90–18526 and ASC 93-18185

and an R.C. Tolman Fellowship awarded to M.B.D.

*Subject headings*: **binaries:general**

**— globular clusters:general:M3 — stars:kinematics**

## 1. Introduction

Since the discovery of BSs in M3 by Sandage (1953), the origin of this anomalous stellar population has been a challenge to astrophysicists. A number of hypotheses have been proposed to explain their origin (see reviews by Nemec 1989, Leonard 1989, Livio 1993, Trimble 1993). Currently favoured scenarios are that the BSs formed from binaries, either by mass–transfer in close primordial binaries (McCrea 1964) or by merger during stellar collisions (Hills & Day 1976). At least some of the globular cluster BSs are descended from mass–transfer binaries (Mateo *et al.* 1990), and the population of primordial binaries formed with semi–major axis small enough that stellar evolution, gravitational radiation or magnetic braking (Iben & Tutkov 1984) will drive mass–transfer in the binary must provide a background rate for BS formation. It has become apparent that the numbers and distribution of BSs in different globular clusters is such that at least one other formation channel must be at work (see eg. Livio 1993).

Recently, formation scenarios for BSs involving stellar mergers during binary interactions have gained favour (Hut & Verbunt 1983, Leonard 1989, Bailyn 1992, Lauzeral *et al.* 1992, Sigurdsson & Phinney 1993 [hereafter SP93], Livio 1993; Davies, Benz and Hills 1994 [hereafter DBH94]). Here we consider the predicted radial distribution of BSs formed via primordial binary interactions, in a moderate concentration globular cluster, such as M3 with density $\rho = 3 \times 10^3$ $M_\odot$ pc$^{-3}$, projected dispersion $\sigma \sim 5.4$ km s$^{-1}$, concentration $c = 2.0$ (Trager *et al.* 1993).

One would expect the production of BSs, via stellar mergers, to be concentrated in the core of a globular cluster, where the stellar number densities are largest. Additionally, the effects of mass–segregation will further increase the central densities of the more massive components of the cluster, including the BSs. However, recent observations of M3 run counter to this expectation; the BSs show a significant underdensity at about $5r_c$ ($r_c$ being the core radius), an excess at $\sim 10r_c$ and a possible underdensity in the core (Bolte *et al.* 1993, Ferraro *et al.* 1993).

## 2. Dynamics of BS

We assume that BSs are formed in the core of the globular cluster by a stellar collision during a binary interaction. The binary is assumed to be primordial, although the orbital parameters and binary membership will likely have been modified from their initial value due to 15 Gyears of interactions that perturbed the binary semi–major axis and eccentricity but did not lead to a stellar collision (Hills 1992, SP93). The presence of "hard" primordial binaries in globular clusters provides an enormously increased cross–section for stellar collisions due to resonant encounters, whether during binary–single (Hut & Bahcall 1983, Hut *et al.* 1992 [and references therein], SP93) or binary–binary (Mikkola 1983, Leonard 1989, Hut *et al.* 1991) interactions. Typically, resonant binary–binary interactions proceed through the prompt ejection of the lightest star in the encounter, with the collision occurring during the resonant interplay of the remaining three stars. As all memory of initial conditions is very rapidly lost during the interaction, the dynamics of the collision during a binary–binary encounter is well represented by binary–single collisions drawn from an appropriately restricted phase space. Only rarely will a binary–binary encounter lead to a quadruple merger, and then mostly for very hard binary interactions (Goodman & Hernquist 1991, Sigurdsson & Hernquist 1992). Quadruple mergers should be vanishingly rare in the moderate density globular cluster we model here, as the contribution to the encounter rate is dominated by binaries wide enough that the fractional cross–section for multiple mergers is small.

Smoothed Particle Hydrodynamics (SPH) simulations of collisions during binary–single interactions (DBH94) show there can be dynamically significant recoil during merger. This is further evidenced by simulations of point mass encounters, covering larger phase space, approximating the collision as an instantaneous inelastic merger (SP93, DBH94). A typical stellar collision will involve the two most massive stars, with a a third "spectator" star ending up in either a wide eccentric orbit about the merged remnant, or escaping to infinity. Occasionally the "spectator" star will collide with the remnant, or its bloated envelope, particularly for low angular momentum encounters involving tight binaries. The orbit of any third, bound "spectator" star may be wide enough that subsequent encounters will rapidly strip it from the BS (SP93, DBH94).

There are two primary mechanisms by which the merged remnant of a stellar collision may receive net momentum in the encounter. SPH simulations show moderate mass–loss during the stellar collision. Any asymmetry in the outflow, as may be expected by grazing incidence collisions, leads to a net kick on the remnant in the centre–of–mass frame. This produces a net kick to the presumptive BS whether or not it remains bound in a (wide) binary. Typical recoils are of order 3 $\rm km\,s^{-1}$, with a tail out to 30 $\rm km\,s^{-1}$ (DBH94). The recoil, on average, adds in quadrature to the net velocity of the encounter centre–of–mass frame relative to the rest frame of the globular cluster. A second kick mechanism comes simply from the separation velocity of the merged remnant and the "spectator" star for those encounters in which the "spectator" star escapes to infinity. The stellar collision can be considered as an inelastic collision superposed on a super-elastic scattering event, with the constituent stars separating with some velocity in the centre–of–mass frame characteristic of the net energy released by the encounter. A simple instance of this is the "Scattering Induced Merger (SIM)" discussed in DBH94. A SIM occurs when a field star abruptly hardens a binary to the point where the periastron of the resulting binary becomes small enough for tidal effects to be significant (or the stars simply collide on the next periastron passage). The resulting release in binding energy leads to significantly larger separation velocity than the encounter velocity in the centre–of–mass frame, or a net kick in the globular cluster rest frame. The distribution of kicks depends on the inelasticity of the collision and on the binding energy released in the effective hardening of the two stars to contact. The distribution of kicks must be found numerically, as the close approach may be due both to a hardening of the binary (semi–major axis, $a$, decreases towards contact) and angular momentum loss (eccentricity, $e$, increases), indeed, during resonant encounters the stars colliding may be on mildly hyperbolic orbits relative to each other, with $e > 1$ at contact. The number of SIMs (and the related Resonant Induced Mergers [RIM]) simulated with SPH is insufficient to provide a good kick distribution, however SPH simulations suggest that the totally inelastic merger approximation for point mass particles may provide an adequate approximation for the dynamics (DBH94). The net kick distribution during stellar collision then is assumed to come from both mechanisms, with the R/SIM distribution drawn from the point mass simulations described in SP93. The resulting kick distribution is approximated by a two component distribution. Given the density and dispersion of M3, the cross-section for stellar collisions is dominated by binaries with semi–major axis, $a$, of order $0.3 - 3 \, AU$. From our simulations, we draw the resulting expected kick distribution due to mass loss and recoil. For this model we assume a primordial binary distribution with a uniform distribution in $\log a$. We infer half the collisions result in the "spectator" star bound to the merged remnant, with a kick distribution uniformly distributed between $1.5 - 10 \, \mathrm{km \, s^{-1}}$, due to mass loss, and with half the collisions resulting in unbound remnants recoiling with a kick distribution uniformly distributed between $0 - 27 \, \mathrm{km \, s^{-1}}$, due to a combination of mass–loss and separation recoil (this allows for a fraction of the bound systems with the "spectator" star in orbit wide enough that it will be stripped from the remnant by encounters with field stars on a time scale short compared to the lifetime of the remnant). This is a reasonable approximation for stellar collisions expected for the range of binaries expected in a cluster with the density and dispersion of M3. The results were not sensitive to the exact shape of the distribution.

For denser clusters, the collision cross-section is dominated by closer binaries, indeed in the densest clusters any binary in the core would have hardened to $a \lesssim 0.1 \, AU$. For these binaries, the fraction of unbound remnants is much smaller (DBH94,SP93) but with the distribution of recoils extending to higher speeds. In low density clusters, the collisions rate is dominated by binaries with larger $a$, with more unbound remnants expected, but lower recoil speeds (SP93). The collision cross-section is somewhat sensitive to the assumed primordial binary distribution for the lowest density clusters. For the high density clusters dynamical evolution of the primordial binary parameters removes most of the sensitivity to the initial distribution in $a$.

The encounter was assumed to have occurred with a center–of–mass velocity relative to the globular cluster rest frame characteristic of the dispersion of binaries in thermal equilibrium with the field stars. The kick was assumed to be randomly oriented with respect to the center–of–mass velocity, drawn randomly from the assumed center–of–mass velocity distribution, and the two were vectorially summed to give a net velocity for the merged remnant in the globular cluster rest frame.

The remnant trajectory was then integrated in a multi–mass King model of M3. We generated a multi–mass model for M3, fit to observational data. The model used a $x = 1.0$ power law IMF (slightly shallower than the canonical Salpeter IMF value of 1.35), with a ZAMS ranging from 0.1 to 15 $M_\odot$. A 10 mass density grid was generated assuming a current turn–off mass of $0.8 M_\odot$ and an evolved population of white dwarfs and neutron stars appropriate to the ZAMS IMF (see Sigurdsson 1991 for detailed description and Sigurdsson & Phinney [in preparation]). The remnant, now treated as a point mass, was assumed to move according to

$$\ddot{\mathbf{r}} = \nabla \Psi(r) + \mathbf{a_{dyf}} + \mathbf{a_{diff}}, \tag{2.1}$$

where $\nabla \Psi(r)$ is the potential gradient due to the mass interior to $r$, $\mathbf{a_{dyf}}$ is the dynamical friction experienced by the binary, and $\mathbf{a_{diff}}$ is the effective acceleration due to scattering by individual stars in the cluster. To calculate $\mathbf{a_{dyf}}$ and $\mathbf{a_{diff}}$, we calculated the diffusion coefficients, $D(\Delta(v_i))$, $D(\Delta v_i \Delta v_j)$ explicitly from the distribution function (Binney and Tremaine, 1987) and hence three independent diffusion coefficients, $D(\Delta v_\parallel)$, $D(\Delta v_\parallel^2)$ and $D(\Delta v_\perp^2)$ (Sigurdsson 1991, Sigurdsson & Phinney [in preparation]). Calculating **r** and **v** at each step, the resulting trajectory was integrated explicitly in the cluster potential. An ensemble of 300 "merged" stars was integrated, with the mergers assumed to have occurred uniformly over the lifetime of the presumptive BS phase, here assumed to be $5 \times 10^9$ years. At the end of the integration, the resulting spatial distribution in the cluster was considered and compared to observations.

## 3. Results

The resulting radial distribution of BS, relative to the surface density of turn–off mass stars (the "normalized BS distribution"), is shown in Figure 1. The "zone of avoidance" at around $5r_c$, with comparable normalized densities interior and exterior to this zone, is qualitatively identical to the data presented in Figure 9 of Ferraro *et al.* (1993) which shows their measured BS normalized distribution for M3. This similarity suggests that stellar collisions during binary interactions are a dominant channel for BS formation in M3. The qualitative explanation for the M3 BS distribution is therefore that in a cluster with the density and velocity distribution profile of M3, binaries which receive kicks that send them beyond $\sim 7r_c$ require long enough to drift back to the center of the cluster that they effectively remain out of energy equilibrium with the rest of the cluster stars. Binaries whose kicks take them to $< 7r_c$ rapidly drift back to the center of the cluster and the compact orbits appropriate to their larger-than-average mass. The model is not good enough to rule out contributions from other channels (such as mass–transfer and coalescence in contact binaries) at the 10-20% level.

Binary collisions also lead to exchanges of turn–off mass stars into binaries and hardening of binaries already containing turn–off mass stars, producing a population which will undergo mass-transfer as the primary evolves off the main–sequence, possibly producing a BS, where the original binary may have been too wide to lead to a BS through mass–transfer, or not contained a star that would have entered the giant branch at this time. BSs produced through binary interactions will consist both of direct collision mergers, and binaries hardened to the point where subsequent mass–transfer is important. The recoil distribution is similar in both cases (Phinney & Sigurdsson 1991). In low density clusters such as M3 this cannot be a dominant mechanism as binaries with $a$ small enough that encounters will harden them to the point where mass–transfer may lead to a BS, provide a small contribution to the total encounter rate. This mechanism is likely to be important in denser clusters.

We can make some predictions as to what would be expected for the BS radial distribution in clusters with different central concentrations than M3. Preliminary simulations of BSs formed by collisions during binary interactions show that in more concentrated, higher dispersion models, the normalized BS distribution is strongly peaked in the center (as expected from the shorter dynamical friction time scale), whereas in low concentration, low dispersion models the BS distribution is flatter due to longer dynamical friction time scales and lower kicks.

The core density of M3 is too low for single-star collisions to be significant contributors to BS formation. In denser clusters, such as 47Tuc or M15, the single-single star collision rate may be comparable to the binary-single star collision rate, depending on the surviving primordial binary fraction. We note that the observed radial distribution of BSs in those clusters may be more concentrated towards the core, although the radial coverage of BS searches is much less complete (Guhathakurta *et al.* 1994, in preparation).

While the published M3 data provide the best observational sample available with regard to the radial distribution of BS over an entire cluster, after the current investigation was begun, new observations have been made which suggest *potential* problems with the Ferraro et al. BS distribution. The first issue is whether the specific frequency of BS in the core of M3 is smaller than at $\sim 15 r_c$. A recent preprint based on Hubble Space Telescope observations (Guhathakurta *et al.* 1994) suggests that Bolte et al 1994 missed some of the BS inside of 1 core radius, and the specific frequency in the center found by Bolte et al. may have been too low. Theoretically, the core density of BS is set by the ratio of the BS lifetime to the half–mass relaxation time, both of which are uncertain to within a factor of two, and, to a lesser extent, by the velocity distribution of BS kicks in the $10 - 20\,\mathrm{km\,s^{-1}}$ range. The actual ratio of BS in the core and beyond the half–mass radius will provide interesting constraints on both the structure of M3 and the true velocity distribution of BSs at formation.

There may also be some questions regarding of the "zone of avoidance" reported by Ferraro *et al.* 1993. Bolte (1994) finds approximately twice the number of BS between $90''$ and $150''$ from the cluster core as Ferraro et al. tabulated. If these stars are added to the Ferraro et al. sample, the "zone of avoidance" is still evident in their Fig. 9, although the magnitude and statistical significance of the dip is reduced. These apparent contradictions in the observations can be taken as an indication of the difficulty of accurately counting BS in regions of dramatically different crowding conditions. We believe that the safest procedure is to put the greatest trust into studies which have at least internal consistency as appears to be the case for the Ferraro et al. study between 60 and $\sim 250$ arcseconds from the cluster core. The different BS counts from different studies also indicate the need for extensive tests by the observers to determine the completeness of their samples at different radii and possibly the need to explicitly determine the contamination in the "BS region" of the CMD due to photometric errors in measuring stars from the single-star sequences.

Should further data show a very different radial density profile for the BSs in M3, our model has failed. If the BS trace the light profile, the BSs cannot be more massive than the main–sequence turnoff, and mergers are not a dominant channel for BS formation. A low core density of BS indicates the BS lifetime is comparable to or shorter than the dynamical friction time scale at the radii reached by ejected BS. An overdensity of BS at large radii, most clearly seen in the current data, strongly suggests that BS receive significant kicks at formation, as predicted by the binary merger scenario. An overdensity of BS in the core suggests either the dynamical friction time scale is short compared to BS lifetimes, or that the BS do not receive significant kicks on formation.

Assuming that the formation mechanisms for BS are common to the different galactic globular clusters, complete observations of a range of classes of globulars will strongly discriminate between BS formation mechanisms. Over the diverse range of clusters in the Galaxy we appear to cover the full range of BS origin possibilities. Observationally, it is important to get complete counts of BS as a function of radius in clusters of widely differing central concentration. Also important is to obtain additional information on the genesis of different individual BSs in these clusters, for example by looking for spectroscopic binaries (Mateo *et al.* 1990) or signs of mixing or rapid rotation (Bailyn 1992), in order to estimate better the relative contribution of other formation channels in different concentration clusters.

We would like to thank Piet Hut and Sterl Phinney for helpful discussion. This research was supported in part by NASA grant NAGW-2422, the NSF under Grants AST 90–18526 and ASC 93-18185 and an R.C. Tolman Fellowship awarded to M.B.D.

**Figure Captions**

**Figure 1** Radial distribution of BSs in M3. The figure shows the relative surface density, $\Sigma(r)$, of BSs in M3, it should be compared with Figure 9 in Ferraro *et al.* 1993. $r_c$ is the model core radius. The surface density of the merged stars formed in the core and evolved for $5 \times f \times 10^9$ years in the M3 model is shown *relative to* the surface density of the turn–off branch stars. The normalization is arbitrary, the error bars show the standard error due to Poisson noise in each bin, systematic errors due to the choice of model are not shown. There is a significant decrease in the relative surface density around 4–6 $r/r_c$, as observed. The dip in $\Sigma$ at the smallest radial bin is real, it is sensitive to the choice of the low velocity tail of the kick distribution and therefore suspect. Ferraro *et al.* do not show this dip, but Bolte *et al.* indicate there is a real dip in the relative density of BSs in the core of M3. A few percent of the presumptive BSs remained at the edge of the cluster, at radii larger than plotted here. The distribution was not statistically significant but indicative that a number of BSs should be observed near the edges of globular clusters.

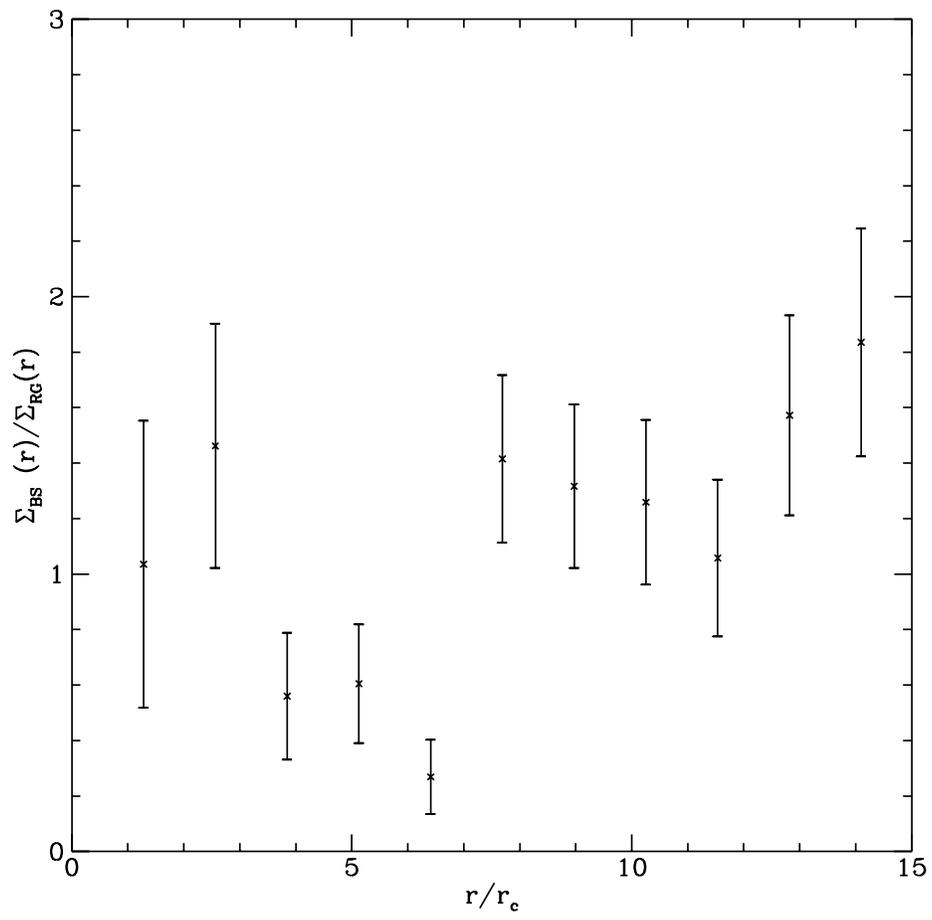